\documentclass[%
 reprint,
 amsmath,amssymb,
 aps,
]{revtex4-2}

\usepackage{graphicx}
\usepackage{dcolumn}
\usepackage{bm}
\usepackage[dvipsnames]{xcolor}
\def\ket#1{ | #1 \rangle }
\def\bra#1{ \langle #1 | }

\begin{document}

\preprint{APS/123-QED}

\title{Quantum paramagnetic 
states in the spin-1/2 distorted honeycomb-lattice Heisenberg antiferromagnet - application to Cu$_2$(pymca)$_3$(ClO$_4$) - }

\author{Tokuro Shimokawa}
 \altaffiliation[]{tokuro.shimokawa@oist.jp}
\affiliation{%
Theory of Quantum Matter Unit, Okinawa Institute of Science and Technology Graduate University, Onna, 904-0495, Japan
}%

\author{Ken'ichi Takano}
\affiliation{
Toyota Technological Institute, Tenpaku-ku, Nagoya 468-8511, Japan
}%

\author{Zentaro Honda}
\affiliation{
Graduate School of Science and Engineering, Saitama University, Sakura-ku, Saitama 338-8570, Japan
}%

\author{Akira Okutani}\altaffiliation[Presnt address:]{Nanophoton Corporation, Semba-nishi, Mino, Osaka 562-0036, Japan} 
\author{Masayuki Hagiwara}
\affiliation{Center for Advanced High Magnetic Field Science, Graduate School of Science, Osaka University, Toyonaka, Osaka 560-0043, Japan
}%




\date{\today}

\begin{abstract}
We investigate the ground-state phase diagram of a spin-1/2 honeycomb-lattice antiferromagnetic (AF) Heisenberg model with three exchange interactions, $J_{\rm A}$, $J_{\rm B}$, and $J_{\rm C}$ that is realized in a distorted honeycomb-lattice antiferromagnet ${\rm Cu_2 (pymca)_3 (ClO_4)}$. We remeasured the magnetic susceptibility of its polycrystalline sample with special care, and determined the exchange parameters of this material through the
comparison with numerical results based on a quantum Monte Carlo (QMC) method. The QMC method also provides a ground-state phase diagram in the $J_{\rm A}/J_{\rm C}$-$J_{\rm B}/J_{\rm C}$ plane.  The phase diagram consists of a small N${\rm \acute{e}}$el phase and a gapped quantum paramagnetic phase surrounding the N${\rm \acute{e}}$el phase.  The latter includes six regimes of hexagonal-singlet-type states and dimer-singlet-type states alternatingly without boundaries closing the spin gap. We further calculate the equal-time spin structure factor in each phase using the QMC method. The computed spin dynamics by the exact diagonalization method exhibits continuums near and in the AF phase. Characteristic four energy band structures in the state with strong hexagonal-singlet-type correlations are informative to clarify the ground-state of ${\rm Cu_2 (pymca)_3 (ClO_4)}$ by future neutron scattering measurements.

\end{abstract}

\maketitle


\section{Introduction}
Understanding the fundamental nature of the quantum magnetism has been an important subject of much investigation, especially by cooperation of experiment and theory. It is easy to understand the importance in the study on quantum frustrated magnets because the combination of the competing interactions and the quantum fluctuations tend to exhibit exotic states of matter such as valence bond crystal (VBC)~\cite{read89,zhitomirsky96,lauchli02,fouet03,albu11}, spin nematic~\cite{blume69,chen71,chandra91,chubkov91,vekua07,kecke07,hikihara08,nic06}, and spin liquid~\cite{anderson73-MatResBull8, balents10-Nature464, savary16, knolle19} states so far, and some of these states are attractive enough for the developments of the future nano devices and quantum computers. Even in the absence of the frustration effect, quantum magnets have much potential to exhibit intriguing phenomena originating from strong quantum fluctuation effects. The Tomonaga-Luttinger liquid~\cite{giamarchi2004quantum} and Haldane~\cite{haldane83,affleck88} states are typical examples in quantum 1D magnets. Generally the role of the quantum fluctuations tends to be less significant in higher dimensional quantum magnets, but, the effect of the quantum fluctuation is likely to be significant in honeycomb-lattice magnets among them because of its minimum coordination number $z=3$ of exchange bonds. The suppression of the classical N${\rm \acute{e}}$el-type long range order (LRO) in the spin-1/2 honeycomb-lattice antiferromagnet (HLA) is larger than that in the spin-1/2 square-lattice antiferromagnet~\cite{Richter04}. Therefore, small perturbations from further-neighbor interactions~\cite{mattsson94,fouet01,takano06,mulder10,mosadeq11,fabio12,li12,zhang13,gong13,zhu17,ghorbani16,ferrari17,ferrari20}, lattice distortion~\cite{takano06,li10,ghorbani16} and randomness~\cite{uematsu17} could be strong enough to destroy the classical LRO, and easy to change the low-temperature physics of the spin-1/2 HLAs.

\begin{figure*}[t]
  \includegraphics[width=16.0cm]{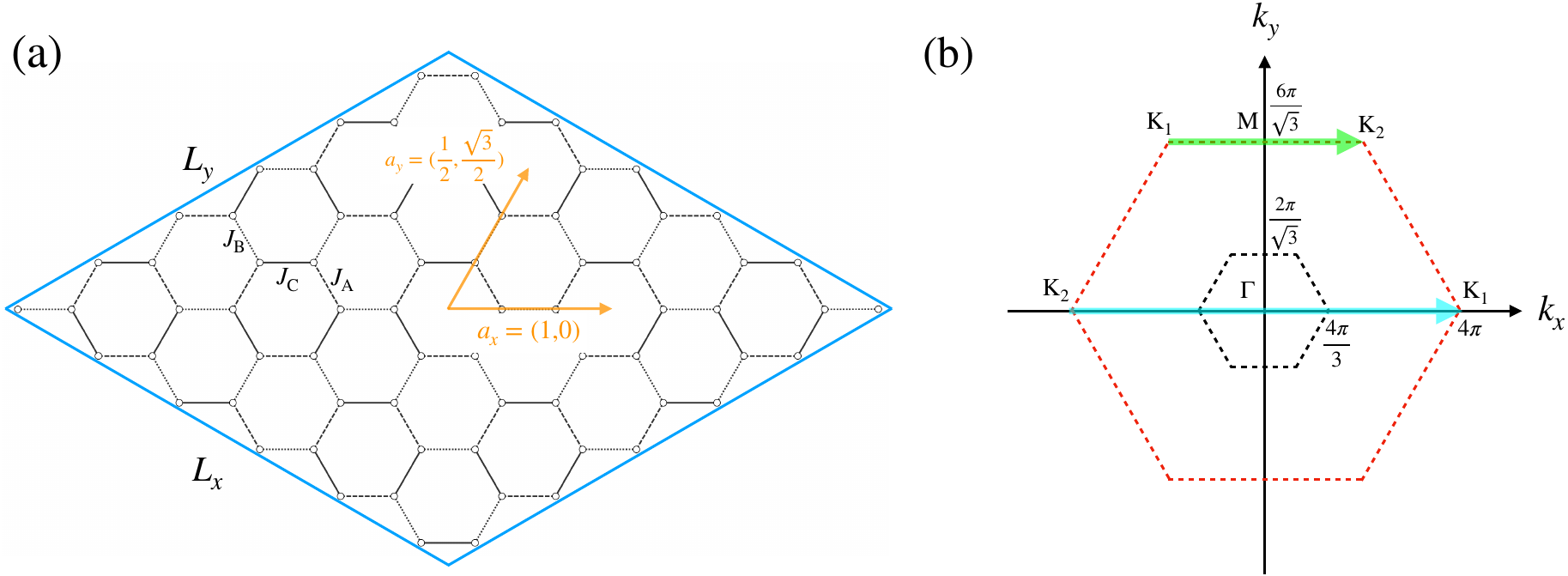}
 \caption{(a) Spin-1/2 $J_{\rm A}$-$J_{\rm B}$-$J_{\rm C}$ honeycomb-lattice Heisenberg model and its finite-size cluster example with $L_x$=$L_y$=6 ($N$=$2 L_x L_y$=72). The orange colored arrows represent the primitive translational vectors in the honeycomb plane. (b) The momentum space obtained by the translational vectors in (a). The black and red dotted lines indicate the Brillouin zone (BZ) for the sublattice triangular-lattice, and the extended BZ for the current honeycomb-lattice, respectively. The green and cyan arrows are used for the horizontal axis in Fig.~\ref{fig07} of the dynamical spin structure factors. }
 \label{fig01}
\end{figure*}

As the candidates of the extended spin-1/2 HLAs, several materials exhibiting nonmagnetic behavior were reported, such as
Na$_3$Cu$_2$SbO$_6$~\cite{smirnova05,miura06,koo08,miura08,kuo12,schmidt13,schmitt14},
Na$_2$Cu$_2$TeO$_6$~\cite{xu05,miura06,shahab07,koo08, schmitt14, gao20},
Li$_3$Cu$_2$SbO$_6$~\cite{skakle97,nalban13,koo16,vavilova21},
Yb$_2$Si$_2$O$_7$~\cite{smolin70,felsche70,nair19,hester19,flynn21}, and
Zn(hfac)$_2$A$_x$B$_{1 - x}$~\cite{yamaguchi17}, where hfac represents 1,1,1,5,5,5-hexafluoroacetylacetonate, and A and B equivalent to regioisomers of verdazyl radical.
The origin of the quantum paramagnetic ground state (GS) in some of the above materials is weakly-coupled ``singlet-dimers''.
For example, the effective model of Na$_3$Cu$_2$SbO$_6$ and Na$_2$Cu$_2$TeO$_6$ is most likely an alternating Heisenberg spin chain with ferromagnetic (FM) and antiferromgnetc (AF) interactions~\cite{schmitt14,gao20}.
The gapped paramagnetic behavior in these two materials could be understood from the spin-1 Haldane chain picture~\cite{haldane83,affleck88,sugimoto20}.
A similar situation also happens in Li$_3$Cu$_2$SbO$_6$, but complicated low-temperature behavior was observed because Li ion acts as a nonmagnetic defect to cut the chains into fragments and generate quasi-free spins at the edge of them~\cite{vavilova21}.
In contrast, the network of the inter and intra dimer couplings form a spatially distorted (breathing) honeycomb-lattice in Yb$_2$Si$_2$O$_7$. The two-dimensionality does not bring this material to undergo a magnetic ordering phase transition at lower temperatures, and the small ratio of the inter to intra dimer couplings, $\sim$0.4, remains this material in the quantum dimer state~\cite{hester19}. It was also reported that the anisotropic perturbations in spin space are required to understand the field-induced successive phase transition in this material~\cite{flynn21}.
Another origin was proposed for a newly synthesized organic compound Zn(hfac)$_2$A$_x$B$_{1 - x}$ exhibiting gapless liquid-like behavior~\cite{yamaguchi17}. The combined effects with randomness in magnetic exchange interactions, frustration, and quantum fluctuation may produce an intriguing random singlet state~\cite{watanabe14,kawamura14,shimokawa15,uematsu17,uematsu18,uematsu19,kawamura19,uematsu21}.


${\rm Cu_2(pymca)_3(ClO_4)}$ was also reported as an extended spin-1/2 HLAs exhibiting gapped paramagnetic behavior down to 2 K~\cite{honda15}, where ``pymca'' is pyrimidine-2-carboxylate. This compound was reported to be a regular honeycomb-lattice copper one with a trigonal symmetry (space group $P31m$), and each pymca ligand connects to two Cu$^{2+}$ ions, forming a honeycomb network in the $ab$ plane~\cite{honda15}. However, the recent single-crystal X-ray diffraction experiments using a synchrotron radiation facility revealed the existence of the distorted octahedral geometry of Cu$^{2+}$, which leads to at least three intralayer exchange interactions in the honeycomb plane~\cite{sugawara17}. 

Additional experiments were conducted on polycrystalline samples of this material~\cite{okutani19}. The measured specific heat at zero-field decreased down to 0.6 K without any anomaly. The electron spin resonance (ESR) spectrum could be fitted to a single Lorentzian function, which indicates the isotropic AF interactions in this material, and the estimated $g$-factor is 2.13. Interestingly, the high-field magnetization measurements up to 70 T showed the zero-field, 1/3, and 2/3 magnetization plateaus and additional plateau-like curvature near saturation. The zero-field magnetization plateau indicates the existence of a quantum-gapped singlet ground state.

The determination of the effective model was also carried out by means of the non-biased quantum Monte Carlo (QMC) method~\cite{okutani19}. The paper~\cite{okutani19} used a Heisenberg model on a distorted honeycomb-lattice, having three different AF interactions, $J_{\rm A}$, $J_{\rm B}$ and $J_{\rm C}$ [see also Fig.~\ref{fig01}(a)], which was reasonably assumed by the results of the X-ray diffraction~\cite{sugawara17} and the high-field magnetization measurements~\cite{okutani19}. The reproduction of the measured magnetization curve was succeeded well except the fourth plateau near saturated field in the case $J_{\rm A}/k_{\rm B}$=$J_{\rm B}/k_{\rm B}$=43.7 K and $J_{\rm C}/J_{\rm A}$=0.2. These calculation results suggest that the effective model of this material is a weakly coupled ``hexagonal singlet'', which is unique from other honeycomb compounds denoted above. The quantum gapped behavior originates from the hexagonal singlet cluster (HSC) in the local hexagon made by the $J_{\rm A}$ and $J_{\rm B}$ interactions~\cite{memo01}. A quite recent theoretical work with triplon analysis and QMC calculations~\cite{adhikary21} also evaluated the experimental exchange couplings which are consistent with those reported in Ref.~\cite{okutani19}, and the relationship between spin-gapped phase and the emergence of the magnetization plateau was discussed based on the obtained ground-state phase diagram of the spin-1/2 $J_{\rm A}$-$J_{\rm B}$-$J_{\rm C}$ honeycomb-lattice Heisenberg model.

In this paper, we first report experimental results of magnetic susceptibility remeasured on newly synthesized polycrystalline powder samples of ${\rm Cu_2(pymca)_3(ClO_4)}$. Combined with large-scale QMC calculations, we succeed in estimating the three exchange parameters, $J_{\rm A}$, $J_{\rm B}$, $J_{\rm C}$ and confirming that the parameters obtained from the magnetization-curve fitting are reasonable. We further investigate numerically the ground-state phase diagram, which was also reported in Ref.~\cite{adhikary21}, but we focus on the parameter dependence of the spin gap value directly calculated by means of the QMC method, which can give useful information about the nature of the quantum gapped paramagnetic states realized in wide parameter region of the spin-1/2 $J_{\rm A}$-$J_{\rm B}$-$J_{\rm C}$ honeycomb-lattice Heisenberg model. We also study the evolution of the zero-field spin dynamics by means of the exact diagonalization (ED) method in the parameter region of $0 \leq J_{\rm A}/J_{\rm C}$=$J_{\rm B}/J_{\rm C}$ $\leq \infty$, including the ``singlet-dimer'', ``AF'' and ``hexagonal singlet'' states. The details of the spin dynamics are informative for understanding the nature of the ${\rm Cu_2(pymca)_3(ClO_4)}$, also of future honeycomb-lattice candidates via inelastic neutron scattering measurements.

The rest of this paper is organized as follows. The experimental details and results on new polycrystalline samples of ${\rm Cu_2(pymca)_3(ClO_4)}$ are first described in Sec.~II. The evaluated three exchange couplings are also presented. Sec. III is devoted to the theoretical part. We provide the basics of our numerical methods and static and dynamical properties of the spin-1/2 $J_{\rm A}$-$J_{\rm B}$-$J_{\rm C}$ honeycomb-lattice Heisenberg model obtained by non-biased QMC and ED methods.
Finally, the paper is summarized with discussion and conclusion in Sec. IV.

\section{Magnetic susceptibility measurement }


Polycrystalline Cu$_2$(pymca)$_3$(ClO$_4$) samples were synthesized by hydrothermal reaction according to the method described in Ref. \cite{honda15}. Magnetic susceptibility $\chi (=M/H$ where $M$ is the magnetization and $H$ is the external magnetic field) of this compound was measured at $\mu_0 H=0.1~{\rm T}$  ($\mu_0$: permeability in vacuum) between the temperature 2 and 300~K using a superconducting quantum-interference device (SQUID) magnetometer (Quantum Design MPMS XL-7). The diamagnetic susceptibility ($-$1.605$\times$10$^{-4}$ emu/f.u. mol) was calculated by using the Pascal's sum rule, and the van Vleck paramagnetic susceptibility (9.552$\times$10$^{-5}$ emu/f.u. mol) was calculated from the $g$-value of Cu$^{2+}$ ion assuming the spin-orbit coupling constant of Cu$^{2+}$ ion in this compound ($-$710 cm$^{-1}$). The observed magnetic susceptibility was obtained by making these corrections. In Ref.~\cite{okutani19}, magnetic susceptibility of a powder sample of ${\rm Cu_2(pymca)_3(ClO_4)}$ was measured after obtaining the magnetization data in high magnetic fields to calibrate the magnetization at low fields using the sample holder for the magnetization measurements. Therefore, the magnetic susceptibility was not accurate and only showed its behavior. This time, we used a well-calibrated capsule as a sample holder for accurate magnetic susceptibility measurements. \par

\begin{figure}[h]
  \includegraphics[width=8.0cm]{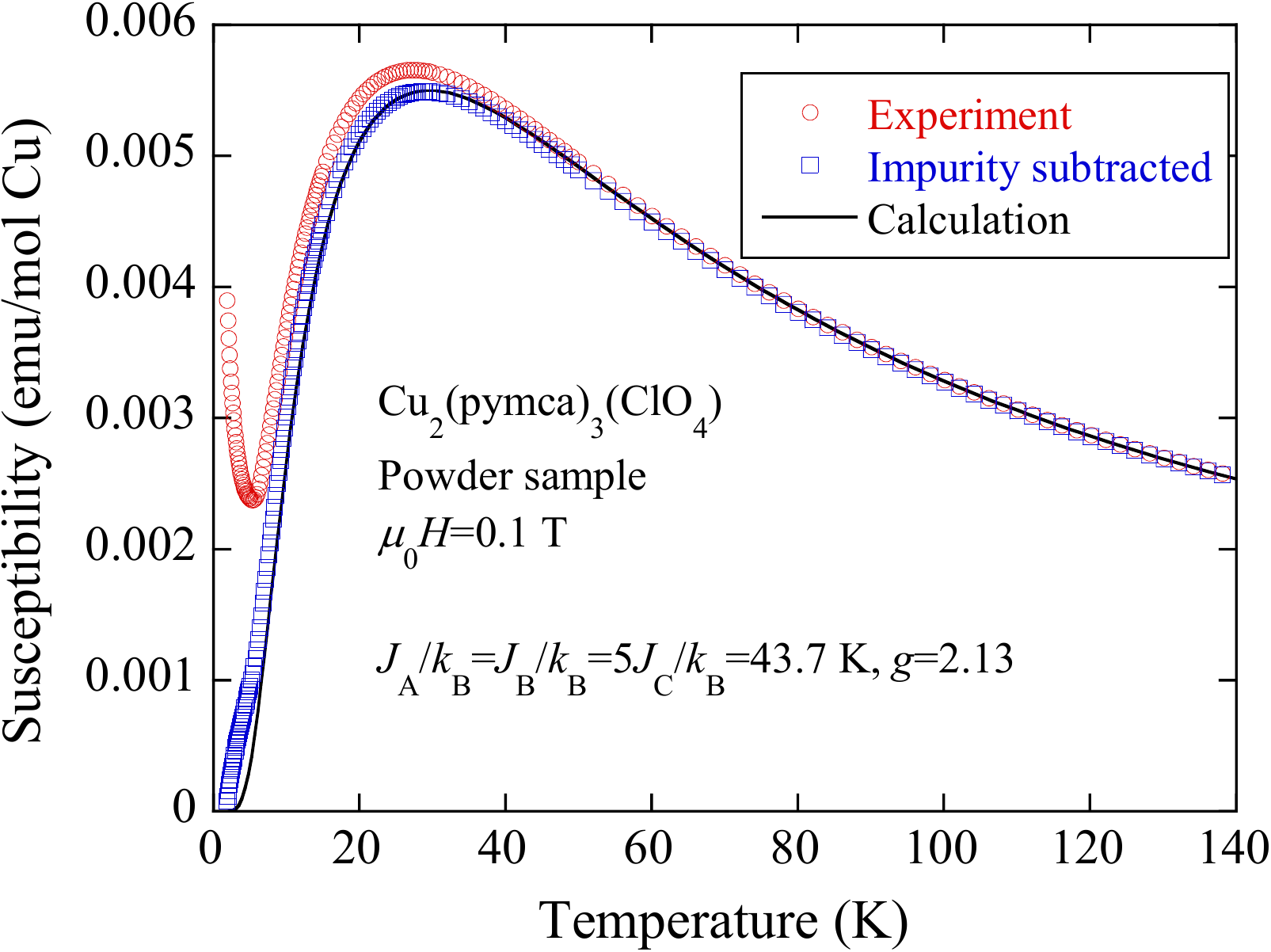}
 \caption{Temperature dependence of the magnetic susceptibility $\chi$ (= $M/H$) of a polycrystalline
${\rm Cu_2(pymca)_3(ClO_4)}$ sample. The open circles represent the measured $\chi$, and the open squares represent $\chi$ minus magnetic-impurity component given by the Curie term. The solid line is the calculated susceptibility using the parameter values given in the figure.}
 \label{fig02}
\end{figure}

The temperature dependence of the magnetic susceptibility of a polycrystalline ${\rm Cu_2(pymca)_3(ClO_4)}$ sample (weight: 72.61 mg)  is shown with open circles in Fig. \ref{fig02}. The magnetic susceptibility shows a broad maximum near 25~K, typical of a low-dimensional antiferromagnet, and a steep increase below 5~K, which might arise from a paramagnetic impurity. To obtain the intrinsic magnetic susceptibility of the sample, we subtracted the paramagnetic-impurity component, given by the Curie term, from the measured magnetic susceptibility (open circles). For this calculation, we assumed the paramagnetic component with $S=1/2$. It is expressed as $\alpha C/T$, where $C$ is the Curie constant with the $g$-value $g = 2.13$ as determined from ESR measurements  at the lowest temperature and the impurity concentration $\alpha = 1.7~\%$. The resultant magnetic susceptibility (open squares) shows a monotonic decrease toward zero upon cooling from 20~K. The calculated susceptibility is given by a solid line using the exchange constants $J_{\rm A}/k_{\rm B}$=$J_{\rm B}/k_{\rm B}$=5$J_{\rm C}/k_{\rm B}$=43.7 K and the $g$-value of 2.13. These parameter values are the same as those used for the magnetization-curve fit in Ref.~\cite{okutani19}. The agreement between the experiment and calculation is satisfactorily good. The peak structure of magnetic susceptibility is well reproduced by the calculation, although its low-temperature part deviates a bit from the calculated one. \par




\section{Numerical study}

\subsubsection{Methods}
Our QMC method is based on the directed loop algorithm in the stochastic series expansion representation~\cite{sandvik99}. The calculations for spin-1/2 $J_{\rm A}$-$J_{\rm B}$-$J_{\rm C}$ honeycomb-lattice Heisenberg model are performed using finite-size clusters $N$=$2 L_x L_y$ (see also Fig.~\ref{fig01}(a)) under the periodic boundary condition, where $N$ is the number of spins. For evaluating the spin gap value of each finite-size cluster, we use the multi-cluster loop algorithm in the continuous-time path-integral representation and the second-moment method. All QMC calculations for magnetic susceptibility, spin gap, and two-point correlation functions are carried out using the ALPS applications~\cite{todo01,ALPS2007,bauer11}.

To compute the zero-temperature dynamical spin structure factor (DSSF), we use the continued fraction expansion~\cite{gagliano87} based on the exact diagonalization (ED) Lanczos algorithm. The equation of the DSSF is written by

\begin{equation}
S^{\alpha}_{\bf q}(\omega) = -\frac{1}{\pi} {\rm Im}\bra{\phi} \hat{S}^{\alpha \dagger}_{\bf q} \frac{1}{\omega - \hat{H} + E_0 + i\eta } \hat{S}^{\alpha}_{\bf q} \ket{\phi},  
\end{equation}
where $E^{~}_0$ is the ground state energy (i.e., lowest eigenvalue) with the corresponding ground state $\ket{\phi}$ of the Hamiltonian $\hat H$ and positive real number $\eta$ is the broadening factor. 
We can rewrite the above equation as 
\begin{equation}
S^{\alpha}_{\bf q}(\omega) = -\frac{1}{\pi} {\rm Im}
\cfrac{\bra{\phi} \hat{S}^{\alpha\dagger}_{\bf q} \hat{S}^{\alpha}_{\bf q} \ket{\phi}}
{z-\alpha^{~}_1-\cfrac{\beta^2_1}
{z-\alpha^{~}_2-\cfrac{\beta^2_2}
{z-\alpha^{~}_3-\cdots
}}}
\label{eq:cfe}
\end{equation}
with $z=\omega - E^{~}_0 + i\eta$. 
${\bm \alpha}$ and ${\bm \beta}$ in Eq.~(\ref{eq:cfe}) are obtained by the tridiagonalization procedure of the Hamiltonian matrix 
in the Lanczos iteration. We here use the cluster of $L_x$=3 and $L_y$=6 ($N$=36), and set $\eta$=0.01 for our DSSF computations. The wave vector points ${\bf q}$ we handle are only along the green and cyan arrows in Fig.\ref{fig01} (b).

\subsubsection{Computational results}
\begin{figure}[t]
  \includegraphics[width=8.0cm]{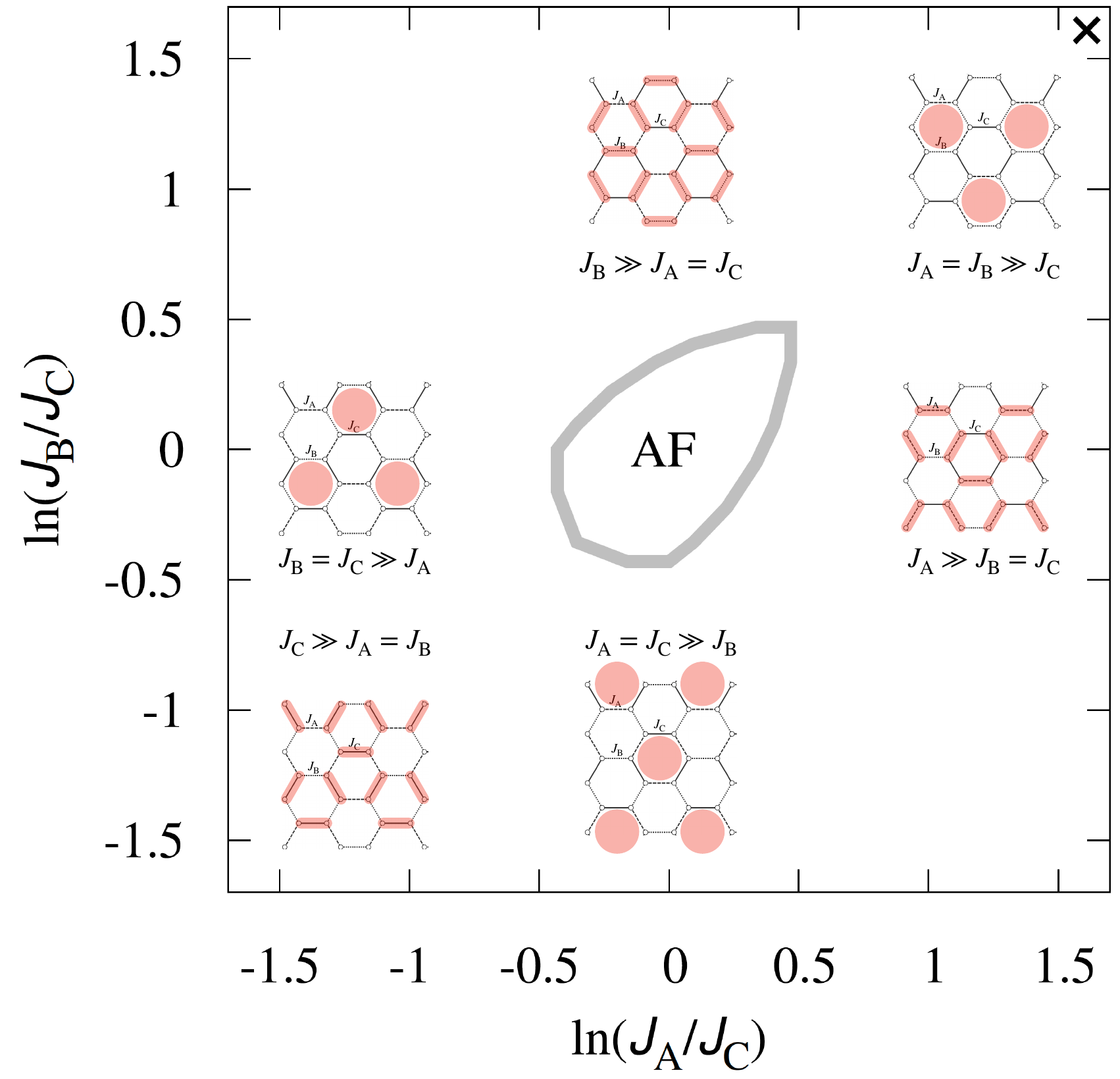}
 \caption{The zero-field ground-state phase diagram obtained by the QMC calculations. The six insets depict the schematic pictures of the real-space correlation pattern in the gapped quantum paramagnetic states. The cross point corresponds to ${\rm Cu_2(pymca)_3(ClO_4)}$ and is located in one of the HS areas}
 \label{fig03}
\end{figure}

We first show our ground-state phase diagram in Fig.~\ref{fig03} evaluated by the spin gap. One can find that the region of the antiferromagnetic N${\rm \acute{e}}$el phase is small and surrounded by the quantum paramagnetic phase. We also show the schematic pictures of the dominant correlation patterns in 6 parameter limits, respectively, like the insets in Fig.~\ref{fig03}. The obtained phase diagram is consistent with that determined by the different QMC calculations using sublattice magnetization and spin stiffness~\cite{adhikary21}. 

\begin{figure*}[t]
  \includegraphics[width=17.0cm]{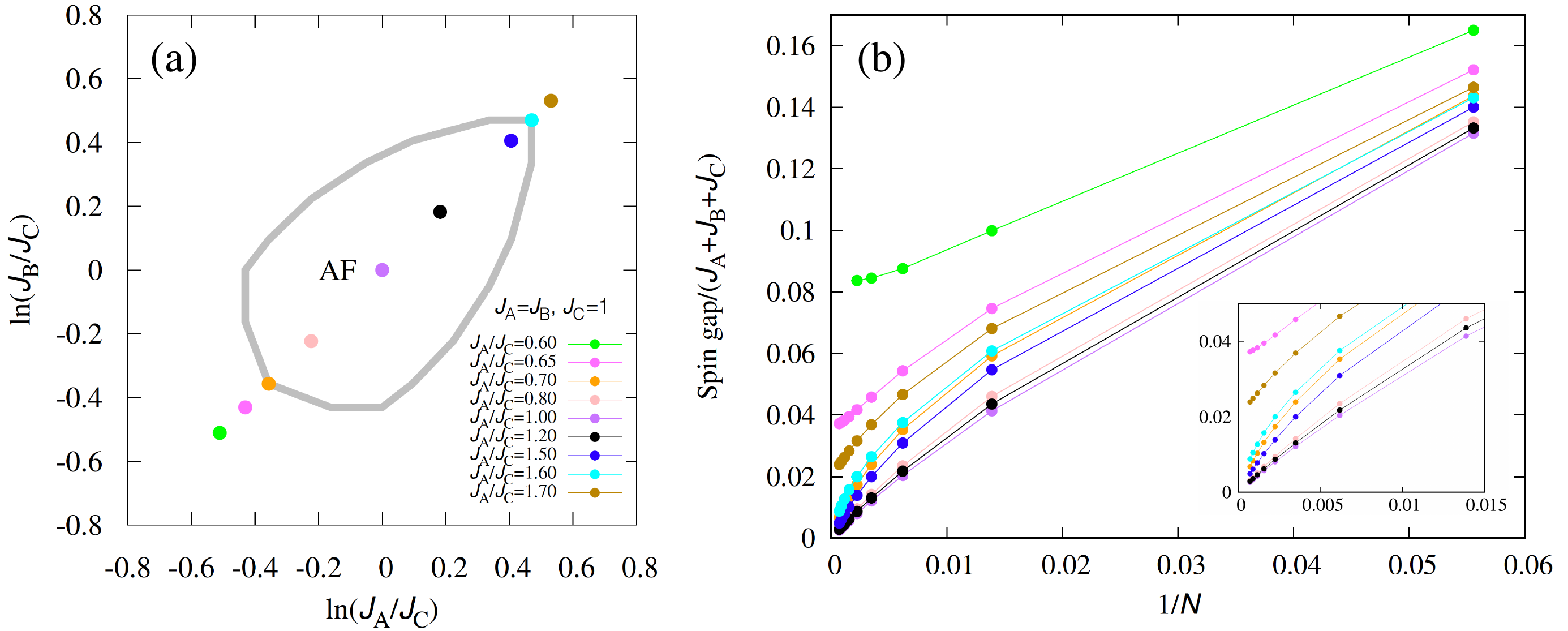}
 \caption{(a) The focused view of the zero-field ground-state phase diagram around the AF phase. Each color point corresponds to the parameter value used in (b). (b) The size dependence of the calculated spin gap using the QMC method. The inset shows the focused view for a larger $N$ region.  }
 \label{fig04}
\end{figure*}

Let us explain the way to determine the phase boundary between quantum paramagnetic and the N${\rm \acute{e}}$el phases. The spin gap calculations are employed on the finite-size clusters of $L_x$=$L_y$=3,~6,~9,~12,~...,~24, and 27. For example, we show the size dependence of the spin gap along the $J_{\rm A}$/$J_{\rm C}$=$J_{\rm B}/J_{\rm C}$ line in Fig.~\ref{fig04}. The treated $J_{\rm A}$=$J_{\rm B}$ values are depicted in Fig.~\ref{fig04}(a). To estimate the spin gap value in the thermodynamic limit, $\Delta_{\infty}$, we use a fitting equation $\Delta_N$=$\Delta_{\infty}+A$${\rm exp}(-B N^{1/2})$, where the $A$ and $B$ are fitting parameters~\cite{nakano11,matsumoto01} when we assume the presence of the finite spin gap. We find that the fitting curves seem to be appropriate for $J_{\rm A}/J_{\rm C}$=$J_{\rm B}/J_{\rm C} \lesssim 0.70$ and $J_{\rm A}/J_{\rm C}$=$J_{\rm B}/J_{\rm C} \gtrsim 1.60$. 

\begin{figure}[b]
  \includegraphics[width=8.0cm]{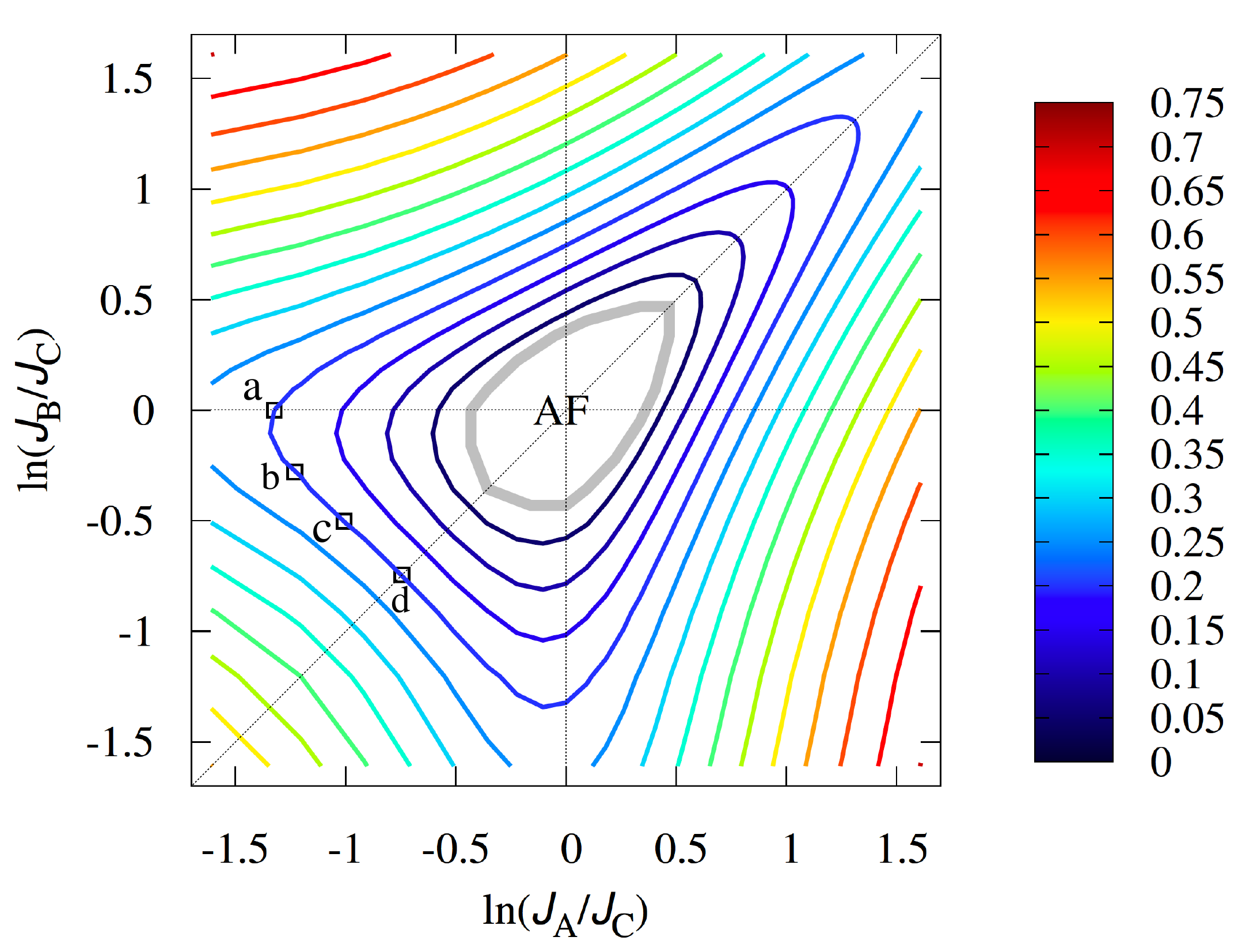}
 \caption{The contour plot of the spin gap value in the finite-size cluster, $L=12$ ($N$=288). The color scale means the value of the spin gap. The labels a-d provide the corresponding parameter values used in Fig.~\ref{fig06}.}
 \label{fig05}
\end{figure}

To further investigate the evolution of the spin gap in the parameter space, we display the contour plot of the spin gap value using $L_x$=$L_y$=12 cluster, in Fig.~\ref{fig05}. The shape of the evolution looks like a deltoid curve and is symmetric with respect to the line of $J_{\rm A}$=$J_{\rm B}$. We also clearly see that the six gapped regimes alternatingly appear without closing the energy gap as depicted in Fig.~\ref{fig03}.

\begin{figure*}[t]
  \includegraphics[width=17.0cm]{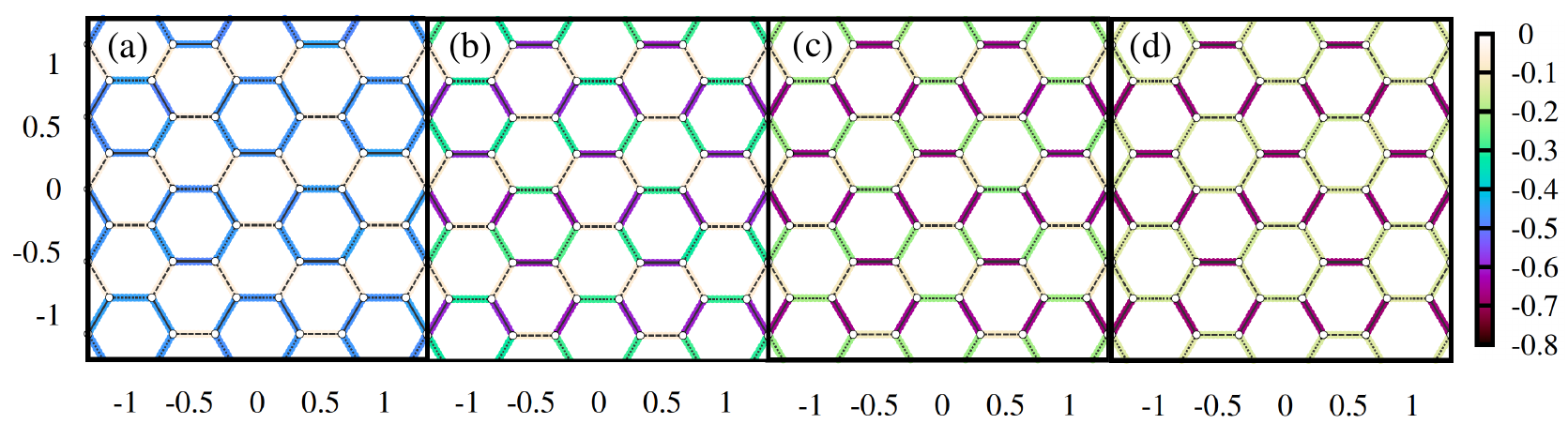}
 \caption{(a-d) The evolution of the real-space nearest-neighbor two-point correlation functions along a line with a constant spin gap. The color on each bond represents the expectation value of the two-point correlation function, $ \langle {\bf S}_i \cdot {\bf S}_j \rangle$, on each $i$-$j$ bond. The exchange parameters used in (a-d) are shown in Fig.~\ref{fig05}. }
 \label{fig06}
\end{figure*}

We compute the nearest-neighbor two-point correlation function to see the cross-over phenomenon in the quantum paramagnetic phase. As a characteristic example, we deal with four-parameter sets of ($J_{\rm A}/J_{\rm C}$, $J_{\rm B}/J_{\rm C}$) having the same constant spin gap, and these four parameter points are shown as the square symbols denoted by a-d in Fig.~\ref{fig05}, respectively. The calculation results on the $L_x$=$L_y$=12 cluster are shown in Fig.~\ref{fig06}. The color on each bond represents the value of nearest-neighbor two-point correlation function, $ \langle {\bf S}_i \cdot {\bf S}_j \rangle$, where $i$ and $j$ are the edge sites of the corresponding bond. From (a) to (d), one can confirm the replacement of the dominant spin correlations are changed from a hexagon-like to a dimer-like pattern.

Note that in Fig.~\ref{fig05}, the hexagonal-singlet state at point a and the dimer-singlet state at point d are on the same contour line, but point a from the center of the AF state ($J_{\rm A}/J_{\rm C}$=$J_{\rm B}/J_{\rm C}$=1) is farther than point d. This means that the development of the spin gap is slower along the direction from the isotropic AF state to a hexagonal-singlet state than from the isotropic AF state to a dimer-singlet state, reflecting the spatial largeness of the hexagonal singlet cluster.

\begin{figure*}[t]
  \includegraphics[width=18cm,angle=0]{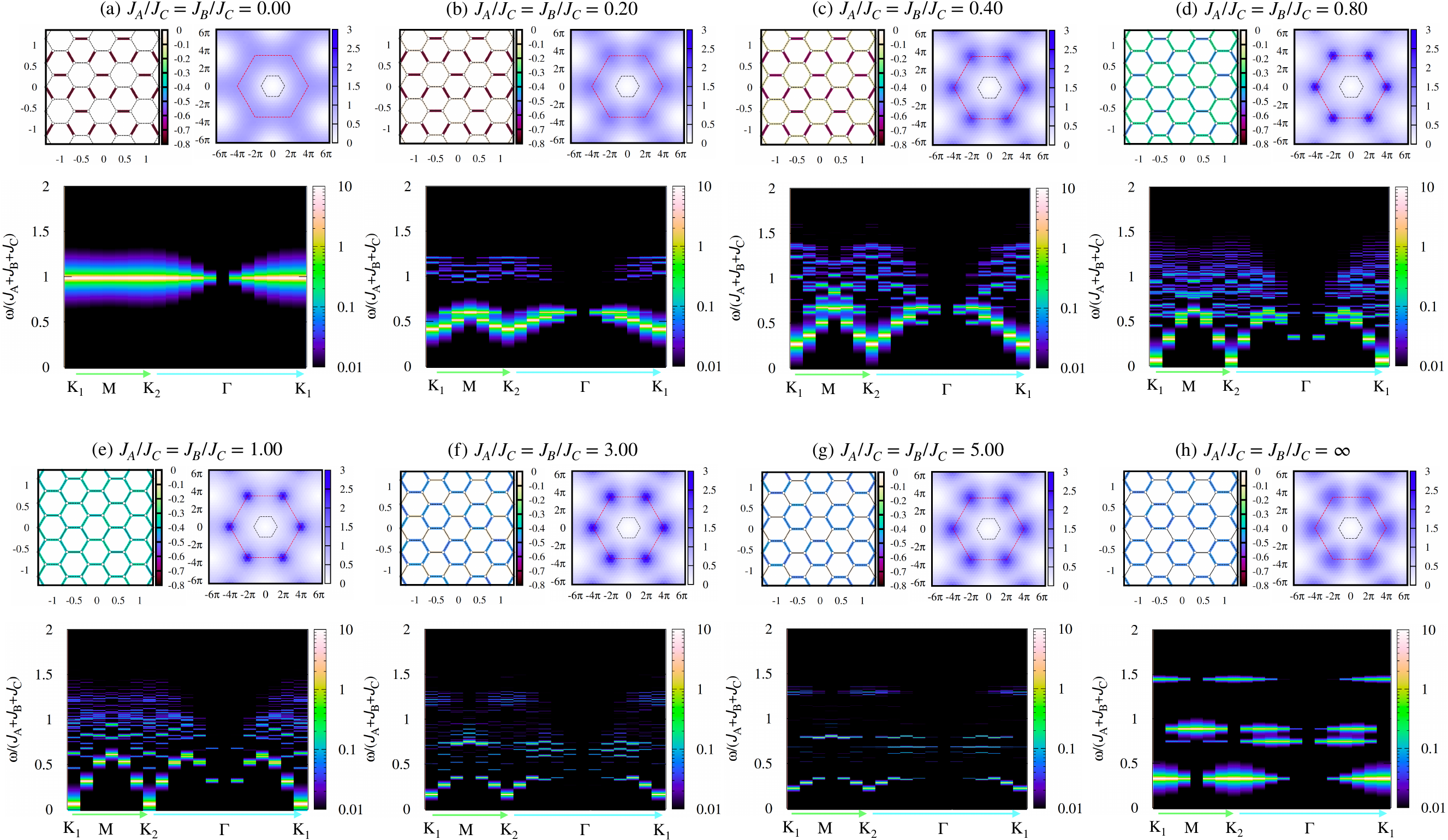}
 \caption{The evolution of the nearest-neighbor spin-spin correlation, the equal-time spin structure factor, and the dynamical spin structure factor along the parameter line with $J_{\rm A}/J_{\rm C}$=$J_{\rm B}/J_{\rm C}$. The former two quantities are calculated using the QMC method based on the finite-size cluster, $L$=12 ($N$=288), and the other one is obtained by the continued fraction expansion employed with the Lanczos method on the finite-size cluster, $L_x$=3 and $L_y$=6 ($N$=36). The momentum zone cut we choose for the dynamical spin structure factors is shown by the green and cyan colored arrows, which are shown in Fig.~\ref{fig01}(b). The results in (a) and (h) correspond to those in the perfectly decoupled singlet-dimer state ($J_{\rm A}$=0) and the hexagonal singlet state ($J_{\rm B}$=$\infty$), respectively. The results in (d) and (e) are obtained in the AF phase. (see also the focused view of the ground-state phase diagram in Fig.~\ref{fig04} (a).)}
 \label{fig07}
\end{figure*}

Next, we investigate the evolution of the spin dynamics and the corresponding equal-time spin structure factor along $0 \leq J_{\rm A}/J_{\rm C}$=$J_{\rm B}/J_{\rm C} \leq \infty$ line. This parameter line includes all of the main states in our target honeycomb-lattice Hamiltonian, the perfect dimer state ($J_{\rm A}$=$J_{\rm B}$=0), the N${\rm \acute{e}}$el state ($J_{\rm A}$=$J_{\rm B}$=1) and the hexagonal singlet state ($J_{\rm A}$=$J_{\rm B}$=$\infty$) and the intermediate states between them. The calculation results are shown in Fig.~\ref{fig07} with the figures of the real-space nearest-neighbor two-point correlation functions. Note that we treat $L_x$=$L_y$=12 cluster to calculate the equal-time spin structure factor and the correlation function using the QMC method. On the other hand, we use the exact diagonalization (ED) method to compute the dynamical spin structure factor using the $N$=36 site cluster. The equal-time spin structure factor is evaluated from the computed two-point correlation functions as

\begin{equation}
S_{\bf q} = \frac{1}{N} [ \langle | \sum_{j} {{\bf S}_j {\rm e}^{i {\bf q}\cdot {\bf R}_j}}| \rangle ],  
\end{equation}
where ${\bf R}_j$ is the position vector at the site $j$.

In comparison to the real-space nearest-neighbor two-point correlation functions, the equal-time spin structure factors are not so distinctive from each other among the quantum paramagnetic states in Fig.~\ref{fig07}(a-c) and (f-h). They just exhibit a broad peak behavior around the edge of the extended BZ (red-colored hexagon), namely, K points in the quantum paramagnetic region. On the other hand, the dynamical spin structure factors are qualitatively different from each other among the all-state we handle in Fig.~\ref{fig07}.

In Fig.~\ref{fig07} (a) and (h), the static and dynamical properties are shown for the perfect dimer state ($J_{\rm A}$=$J_{\rm B}$=0) and hexagonal singlet state ($J_{\rm A}$=$J_{\rm B}$=$\infty$), respectively. 
As we expect from the presence of the localized dimer singlet or the hexagonal singlet, one can confirm flat-band-like features in the computed spin dynamics in these two parameter limits. As away from these two limits, we can see clearly, even in the small size cluster of $N$=36, the flat-band feature is collapsed, the spin gap is decreased, and the magnon dispersion grows around the K points with approaching the N${\rm \acute{e}}$el state in the isotropic parameter case ($J_{\rm A}$=$J_{\rm B}$=$J_{\rm C}$).

It may be worth saying more about the evolution of the spin dynamics from Fig.~\ref{fig07}(a) to \ref{fig07}(b). The single flat band in (a) is split into higher and lower bands in (b) and this higher one seems to have a continuum-like behavior even in the small perturbation to the perfect dimer state. The spin dynamics in Fig.~\ref{fig07}(b) looks like the results of the recent inelastic neutron scattering measurements on a breathing honeycomb-lattice compound, Yb$_2$Si$_2$O$_7$~\cite{hester19}. The zero field inelastic neutron scattering data is shown in Fig.~4(a) in the Ref.~\cite{hester19} and it exhibits a slightly curved low-energy band structure and a fog or continuum-like features. Therefore, the overall features are also captured in our computed spin dynamics in Fig.~\ref{fig07}(b) although we should be careful the difference in the spatial arrangement of the singlet-dimers between Fig.~\ref{fig07}(b) in our paper and Fig.~1(a) in Ref.~\cite{hester19}.

An additional remark is that the continuum evolves near and in the antiferromagnetic N${\rm \acute{e}}$el state. The spin dynamics for the spin-1/2 honeycomb-lattice AF Heisenberg model was already investigated in Ref.~\cite{ferrari20} using a variational Monte Carlo (VMC) technique based upon Gutzwiller-projected fermionic states, and the result also exhibits a similar continuum behavior interpreted by the magnon-magnon interactions. The paper \cite{ferrari20} also pointed out that there is an intermediate energy regime having vanishingly small intensities between the continuum regime and the magnon dispersion of the linear spin wave. Such kind of the intermediate energy blank is also confirmed in our ED calculation in Fig.~\ref{fig07}(e), just above the spin wave dispersion. These consistencies with the results in Ref.~\cite{ferrari20} could support the reliability in our ED spin dynamics with the relatively small finite-size cluster even in the gapless state.

The numerical results in Fig.~\ref{fig07}(g) are for our honeycomb-lattice material, Cu$_2$(pymca)$_3$(ClO$_4$). In this parameter of ($J_{\rm A}/J_{\rm C}$, $J_{\rm B}/J_{\rm C}$)=(5.0, 5.0), we have a quantum gapped paramagnetic state having a strong correlation in each local hexagon made by the $J_{\rm A}$ and $J_{\rm B}$ interactions. This static correlation pattern may be similar to the VBC state in the frustrated $J_1$-$J_2$ honeycomb-lattice AF Heisenberg model. However, the discrepancy in the spin dynamics is clear between the frustrated and unfrustrated cases. The frustration effect from the nearest-neighbor interaction, $J_2$, to the spin-1/2 honeycomb-lattice Heisenberg antiferromagnet gives rise to a significant renormalization of the spin wave magnon dispersion, production of roton-like minimums, and enhancement of the intensities in the broad continuum~\cite{ferrari20}. On the other hand, the hexagonal-type distortion in our honeycomb-lattice model doesn't reproduce them, and there are just four slightly curved energy bands in Fig.~\ref{fig07}(g). This difference could be useful for the future inelastic neutron scattering measurements not only for Cu$_2$(pymca)$_3$(ClO$_4$) but also for other honeycomb-lattice candidates to confirm the presence or absence of the next-nearest neighbor interactions in them.

\section{Summary and discussion}
We measured the magnetic susceptibility of the newly synthesized polycrystalline sample of Cu$_2$(pymca)$_3$(ClO$_4$). Combined with the experimental results, we also employed the non-biased quantum Monte Carlo (QMC) calculations to determine the parameter set in the effective model, the spin-1/2 $J_{\rm A}$-$J_{\rm B}$-$J_{\rm C}$ honeycomb-lattice antiferromagnetic Heisenberg model which was originally proposed by the recent X-ray diffraction~\cite{sugawara17} and the high-field magnetization~\cite{okutani19} measurements. Our careful fitting revealed that $J_{\rm A}/J_{\rm C}$=$J_{\rm B}/J_{\rm C}$=5.0, and $J_A$/$k_B$=43.7 K is the best parameter set for Cu$_2$(pymca)$_3$(ClO$_4$).

We could determine the ground-state phase diagram of the above effective model via spin gap directly calculated by our QMC method. To see the evolution of the spin gap in the antiferromagnetic parameter ($J_{\rm A}/J_{\rm C}$,$J_{\rm B}/J_{\rm C}$) space, we investigated the contour plot of the spin gap. We found that it exhibits a deltoid curve and looks symmetric with respect to the line of $J_{\rm A}$=$J_{\rm B}$. The quantum paramagnetic states appearing in six parameter limits are continuously connected without closing the spin gap, and the small AF N${\rm \acute{e}}$el phase is surrounded by the quantum gapped paramagnetic phase.

Our exact diagonalization (ED) results for the dynamical spin structure factors showed several important and distinct properties for future inelastic neutron scattering measurements not only for the Cu$_2$(pymca)$_3$(ClO$_4$) but also for other honeycomb-lattice materials. The hexagonal singlet state in the unfrustrated systems and the valence bond crystal state in the frustrated systems could be distinguishable in the spin dynamics in contrast to the equal-time spin structure factor.

The remaining question for Cu$_2$(pymca)$_3$(ClO$_4$) is a high-magnetic-field feature. The previous research in Ref.~\cite{okutani19} reported the presence of the magnetic plateaus up to 70 T, and the zero-field, 1/3, and 2/3 magnetization plateaus were reproduced accurately by our current effective model with $J_{\rm A}/J_{\rm C}$=$J_{\rm B}/J_{\rm C}$=5.0. However, the plateau-like curvature near saturation is not reproduced by our current effective model~\cite{okutani19,adhikary21}, which means that we need additional perturbations to the spin-1/2 $J_{\rm A}$-$J_{\rm B}$-$J_{\rm C}$ honeycomb-lattice antiferromagnetic Heisenberg model. One of the possibilities may be the Dzyaloshinskii–Moriya (DM) interaction which is allowed on the Cu-Cu bond from the symmetric viewpoint in Cu$_2$(pymca)$_3$(ClO$_4$). Indeed, the magnetization curve of Cu benzoate with the DM interaction exhibits a bending near the saturation field~\cite{matsuo05}. Experimental studies beyond 70 T are also required associated with the further theoretical studies to clarify the true nature of Cu$_2$(pymca)$_3$(ClO$_4$).

\begin{acknowledgments}
T.~S.~ thanks Synge Todo for his helpful comment and advice to use ALPS QMC applications~\cite{ALPS2007}.
T.~S.~ is supported by the Theory of Quantum Matter Unit of the Okinawa Institute of Science and Technology Graduate University (OIST). 
The experimental work was carried out at the Center for Advanced High Magnetic Field Science in
Osaka University under the Visiting Researcher’s Program of the Institute for Solid State
Physics at The University of Tokyo.
Our numerical calculations were performed using facilities of the Supercomputer Center, ISSP, the University of Tokyo, and using Research Center for Computational Science, Okazaki, and using the computing section, OIST. The work is partly supported by KAKENHI Nos. 19K14665 and 21K03477.
\end{acknowledgments}

\appendix


\nocite{*}

\bibliography{apssamp}

\begin{thebibliography}{77}
\bibitem [{mem()}]{memo01}%
  \BibitemOpen
  \href@noop {} {\bibinfo {title} {{This quantum gapped paramagnetic state may
  have a similarity in the static correlation pattern with that in the valence
  bond crystal (VBC) state proposed in the spin-1/2 $J_1$-$J_2$
  honeycomb-lattice Heisenberg
  antiferromagnet~\cite{mosadeq11,gong13,zhu17,ferrari17,ferrari20}. However,
  the spontaneous translational symmetry breaking occurs in the VBC state, and
  this discrete symmetry could be broken at a finite temperature associated
  with a sharp anomaly in thermodynamic quantities such as specific heat. On
  the other hand, the current gapped paramagnetic state is directly connected
  from a high-temperature paramagnetic state without any anomaly, and no 3-hold
  degeneracies in the ground state. In this sense, the current quantum gapped
  paramagnetic state is supposed to be distinct from the VBC
  state.}}}\BibitemShut {Stop}%
\end{thebibliography}

\end{document}